\documentclass[aps,twocolumn,prb]{revtex4-2}
\usepackage{graphicx}
\usepackage{amsmath}

\begin{document}

\title{Mode conversion of extraordinary waves in stratified plasmas with an external
magnetic field perpendicular to the directions of inhomogeneity and wave propagation}

\author{Seulong \surname{Kim}}
\affiliation{Department of Physics, Ajou University, Suwon 16499, Korea}
\author{Kihong \surname{Kim}}
\email{khkim@ajou.ac.kr}
\affiliation{Department of Physics, Ajou University, Suwon 16499, Korea}

\begin{abstract}
We study theoretically the mode conversion and the resonant absorption of high frequency electromagnetic waves into longitudinal modes in magnetized and stratified plasmas
in the case where the external magnetic field is perpendicular to both the directions of inhomogeneity and wave propagation. Mode conversion is shown to occur only when the waves
are extraordinary waves. We develop an efficient method for calculating the mode conversion coefficient for arbitrary spatial configurations of the plasma density and the external
magnetic field in a numerically exact manner using the invariant imbedding method. We calculate the mode conversion coefficient extensively
as a function of the incident angle, the external magnetic field, and the plasma density in the incident region.
We show that there is strong asymmetry under the sign change of the incident angle and the external magnetic
field and find that the mode conversion coefficient is close to one in certain ranges of the parameter values.
We discuss the implications of our results in plasma heating phenomena.
\end{abstract}

\maketitle

\section{Introduction}
\label{sec-intro}

Mode conversion in inhomogeneous plasmas refers to the phenomenon that the energy of transverse electromagnetic waves
is converted into that of local longitudinal modes at various plasma resonances \cite{swanson}.
Since mode conversion plays a crucial role in a wide range of processes occurring in laboratory and space plasmas, it has been studied extensively for many decades \cite{fors,woo,woo2,maki,window2,mjol,hink1,yin,golda,kskim,kim4,kk2,ehk,hood,lee2,ehk2,20,21,22,23,yu}.
The simplest case is the conversion of electromagnetic waves to local plasma oscillations in cold unmagnetized plasmas. The efficiency of the mode conversion is determined by the
strength of coupling between the transverse and longitudinal modes and depends on the form of inhomogeneity and the direction of wave propagation. In magnetized and stratified plasmas,
the relative direction of the external magnetic field with respect to those of inhomogeneity and wave propagation plays an important role in determining the mode conversion coefficient,
which is defined as the fraction of the incident wave energy converted into the local oscillation energy. Since there are three independent directions in the problem, the calculation of
the mode conversion coefficient in the most general case is highly complicated. Therefore the mode conversion in magnetized plasmas has been studied primarily in simplified cases where
the external magnetic field is either parallel or perpendicular to the direction of inhomogeneity.

In \cite{kk2}, the authors have studied the mode conversion in magnetized plasmas
in the case where the external magnetic field is perpendicular to the direction of inhomogeneity and parallel to
the plane of wave propagation. In this paper, we consider another configuration where the external magnetic field is perpendicular to both the direction of inhomogeneity
and the plane of wave propagation. In this case, which has been considered previously in \cite{woo,woo2,maki}, the ordinary (O) and extraordinary (X) wave modes are completely
decoupled. In the case of O waves, no resonance exists and mode conversion does not occur, while, in the case of X waves,
mode conversion occurs if there exists a region where the upper hybrid resonance condition, $\omega^2={\omega_p}^2+{\omega_c}^2$, in which $\omega$, $\omega_p$, and
$\omega_c$ are the wave frequency, the electron plasma frequency, and the electron cyclotron frequency, respectively, is satisfied inside the inhomogeneous plasma.
In the present work, we develop an efficient method for calculating the mode conversion coefficient
in arbitrary spatial configurations of the plasma density and the external magnetic field
in a numerically exact manner using the invariant imbedding method \cite{bell,gol,rammal,kim1,kly1,sk}.
Since the wave equations exhibiting mode conversion contain a singularity at the resonance point, it is difficult to solve them accurately and a variety of approximate methods
have been used in previous researches. Therefore the invariant imbedding equations derived in this paper, which allow an exact numerical solution
of the wave equations with singularities,
will be highly valuable in the study of mode conversion phenomena.

We perform an extensive calculation of the mode conversion coefficient
as a function of the incident angle, the external magnetic field, and the plasma density in the incident region
and compare the results with those of previous works.
We identify the parameter region where the mode conversion is almost perfect. We also
clarify the asymmetry of physical quantities for the changes of the incident angle from $\theta$ to $-\theta$
and of the direction of the external magnetic field to the opposite direction.
The results presented in this paper are expected to be useful for further studies of mode conversion and resonant absorption in inhomogeneous plasmas
and nonreciprocal wave propagation
in media without time-reversal symmetry.

The rest of this paper is organized as follows.
In Sec.~\ref{sec-wave}, we derive the electromagnetic wave equations in cold magnetized plasmas. In Sec.~\ref{sec-iie}, the invariant imbedding equations are derived
using the method described in App.~\ref{sec-imbed}.
In Sec.~\ref{sec-conf}, we explain the plasma density profile used in our calculations. In Sec.~\ref{sec-num}, our numerical results obtained using the invariant imbedding method
are described in detail. Finally, in Sec.\ref{sec-con}, we conclude the paper with some discussions.

\section{Wave equations}
\label{sec-wave}

We consider the high frequency electromagnetic wave propagation and
the mode conversion in stratified and magnetized plasmas. In this paper, we restrict our interest to the case where the
external magnetic field is applied perpendicularly both to the direction
of inhomogeneity and to the wave propagation plane.
We assume that the plasma density varies only in the $z$ direction
and the uniform external magnetic field ${\bf B}_0$ ($=B_0\hat{\bf
y}$) is directed in the $y$ direction. The cold plasma dielectric
tensor, $\epsilon$, for high frequency electromagnetic waves in the present geometry
is written as
\begin{equation}
\epsilon=\begin{pmatrix} \epsilon_1 &0&-i\epsilon_2\cr 0&\epsilon_3& 0\\ i\epsilon_2&
0&\epsilon_1\end{pmatrix}, \label{eq:dtensor}
\end{equation}
where
\begin{eqnarray}
\epsilon_1&=&1-\frac{{\omega_p}^2\left(\omega+i\nu\right)}
{\omega\left[\left(\omega+i\nu\right)^2-{\omega_c}^2\right]},\nonumber\\
\epsilon_2&=&\frac{{\omega_p}^2\omega_c}
{\omega\left[\left(\omega+i\nu\right)^2-{\omega_c}^2\right]},\nonumber\\
\epsilon_3&=&1-\frac{{\omega_p}^2}{\omega\left(\omega+i\nu\right)}.
\end{eqnarray}
We point out that the indices 1, 2, and 3 in $\epsilon_1$, $\epsilon_2$,
and $\epsilon_3$ have no direct relationship with the spatial
coordinates $x$, $y$, and $z$. The constant $\nu$ is the
phenomenological collision frequency. In Gaussian units, $\omega_p$ and $\omega_c$ are given by
\begin{equation}
{\omega_p}^2=\frac{4\pi e^2}{m_e}n(z),~~~\omega_c=\frac{eB_0}{m_e c},
\end{equation}
where $m_e$ and $-e$ are the electron mass and charge respectively.
The electron number density is denoted by the $z$-dependent function $n(z)$.

For monochromatic waves of frequency $\omega$, the wave equations
satisfied by the electric and magnetic fields in cold magnetized
plasmas take the form
\begin{eqnarray}
&&-\nabla\times\left(\nabla\times{\bf E}\right)+{k_0}^2\epsilon\cdot{\bf E}=0,\nonumber\\
&& -\nabla\times\left(\epsilon^{-1}\cdot\nabla\times{\bf
B}\right)+{k_0}^2{\bf B}=0,\label{eq:cwe}
\end{eqnarray}
where $k_0=\omega/c$.
In this paper, we consider only the cases where plane
waves propagate parallel to the $xz$ plane. We assume that all field
components depend on $x$ and $t$ through a factor
$\exp[i(q x-\omega t)]$, where $q$ is
the $x$ component of the wave vector, and have no $y$ dependence. Then it is easy
to derive the wave equations satisfied by the $z$-dependent complex amplitudes $E_y=E_y(z)$ and $B_y=B_y(z)$
from Eq.~(\ref{eq:cwe}),
which take the form
\begin{widetext}
\begin{eqnarray}
&&\frac{d^2 E_y}{dz^2}+\left({k_0}^2\epsilon_3-q^2\right)E_y=0,\label{eq:eo}\\
&&\frac{d}{dz}\left(\frac{\epsilon_1}{{\epsilon_1}^2-{\epsilon_2}^2}\frac{dB_y}{dz}+\frac{q\epsilon_2}{{\epsilon_1}^2-{\epsilon_2}^2}B_y\right)
-\frac{q\epsilon_2}{\epsilon_1}\left(\frac{\epsilon_1}{{\epsilon_1}^2-{\epsilon_2}^2}\frac{dB_y}{dz}+\frac{q\epsilon_2}{{\epsilon_1}^2-{\epsilon_2}^2}B_y\right)
+\left({k_0}^2-\frac{q^2}{\epsilon_1}\right)B_y=0.\label{eq:ex}
\end{eqnarray}
\end{widetext}
The quantities $\epsilon_1$, $\epsilon_2$, and $\epsilon_3$ depend on $z$ through the dependence of $\omega_p$ on $n(z)$.
Once we obtain $E_y$ and $B_y$ by solving Eqs.~(\ref{eq:eo}) and (\ref{eq:ex}), we can calculate other field components using the relationships
\begin{eqnarray}
&& E_x= -\frac{i}{k_0}\frac{\epsilon_1}{{\epsilon_1}^2-{\epsilon_2}^2}\frac{dB_y}{dz}-\frac{i}{k_0}\frac{q\epsilon_2}{{\epsilon_1}^2-{\epsilon_2}^2}B_y,  \nonumber\\
&& E_z= -\frac{1}{k_0}\frac{\epsilon_2}{{\epsilon_1}^2-{\epsilon_2}^2}\frac{dB_y}{dz}-\frac{1}{k_0}\frac{q\epsilon_1}{{\epsilon_1}^2-{\epsilon_2}^2}B_y,  \nonumber\\
&& B_x= \frac{i}{k_0}\frac{dE_y}{dz},  ~~B_z= \frac{q}{k_0}E_y.
\end{eqnarray}

We note that the equations for $E_y$ and $B_y$ are decoupled from each other in the present geometry.
The equation for $E_y$ describes the O wave polarized in the direction of the external magnetic field,
whereas that for $B_y$ describes the X wave polarized in the direction perpendicular to the external magnetic field.
No mode conversion occurs in the O wave case and therefore we consider only the X wave case described by Eq.~(\ref{eq:ex}).
We also point out that the same wave equations are applied to the cases where the external magnetic field $B_0$ is not a constant but a function of $z$.

\section{Invariant imbedding equations}
\label{sec-iie}

We rewrite Eq.~(\ref{eq:ex}) in terms of two coupled first-order differential equations.
We define
\begin{eqnarray}
&&u_1=B_y,\nonumber\\
&&u_2=\frac{\epsilon_1}{{\epsilon_1}^2-{\epsilon_2}^2}\frac{dB_y}{dz}+\frac{q\epsilon_2}{{\epsilon_1}^2-{\epsilon_2}^2}B_y.
\end{eqnarray}
Then we obtain
\begin{eqnarray}
\frac{d}{dz}\left(\begin{array}{cc}
u_1 \\ u_2 \\
\end{array}\right)=
A
\left(\begin{array}{cc}
u_1 \\ u_2 \\
\end{array}\right),\label{eq:meq}
\end{eqnarray}
where the $2\times 2$ matrix coefficient $A$ is given by
\begin{eqnarray}
A=\left(\begin{array}{cc}
-\frac{q\epsilon_2}{\epsilon_1} & \frac{{\epsilon_1}^2-{\epsilon_2}^2}{\epsilon_1}\\
\frac{q^2}{\epsilon_1}-{k_0}^2 & \frac{q\epsilon_2}{\epsilon_1}\\
\end{array}\right).
\label{eq:da}
\end{eqnarray}
We are interested in calculating the reflection and transmission coefficients $r$ and $t$.
A plane wave is incident from the region where $z>L$ and
transmitted to the region where $z<0$. The wave functions in the incident and transmitted regions are expressed in
terms of $r$ and $t$:
\begin{eqnarray}
u_1(z)=B_y(z)=\left\{ \begin{array}{ll} e^{ip(L-z)} +e^{ip(z-L)}~r,
&~z>L\\
e^{-ip^\prime z}~t, &~z<0 \end{array} \right., \nonumber \\\label{eq:psi}
\end{eqnarray}
where $p$ and $p^\prime$ are the negative $z$ components of the wave vector
in the incident and transmitted regions, respectively.

Starting from Eq.~(\ref{eq:meq}) and using the invariant imbedding method described in App.~\ref{sec-imbed}, we derive the invariant imbedding equations for $r$ and $t$:
\begin{widetext}
\begin{eqnarray}
&&\frac{dr}{dl}=2q\left(\frac{{\epsilon_{1}}^2-{\epsilon_{2}}^2}{\epsilon_1}s_2-\frac{\epsilon_2}{\epsilon_1}\right)
+2\left[\frac{{\epsilon_{1}}^2-{\epsilon_{2}}^2}{\epsilon_1}\left(ips_1+qs_2\right)-q\frac{\epsilon_2}{\epsilon_1}\right]r(l)\nonumber\\
&&~~~~+\frac{1}{2ips_1}\left\{-\left(ips_1+qs_2\right)\left[
\frac{{\epsilon_{1}}^2-{\epsilon_{2}}^2}{\epsilon_1}\left(ips_1+qs_2\right)-q\frac{\epsilon_2}{\epsilon_1}\right] -{k_0}^2+\frac{q^2}{\epsilon_1}+q\frac{\epsilon_2}{\epsilon_1}\left(ips_1+qs_2\right)\right\}\left[1+r(l)\right]^2,\nonumber\\
&&\frac{dt}{dl}=\left[\frac{{\epsilon_{1}}^2
-{\epsilon_{2}}^2}{\epsilon_1}\left(ips_1+qs_2\right)-q\frac{\epsilon_2}{\epsilon_1}\right]t(l)\nonumber\\
&&~~~~+\frac{1}{2ips_1}\left\{-\left(ips_1+qs_2\right)\left[
\frac{{\epsilon_{1}}^2-{\epsilon_{2}}^2}{\epsilon_1}\left(ips_1+qs_2\right)-q\frac{\epsilon_2}{\epsilon_1}\right] -{k_0}^2+\frac{q^2}{\epsilon_1}+q\frac{\epsilon_2}{\epsilon_1}\left(ips_1+qs_2\right)\right\}\left[1+r(l)\right]t(l),
\label{eq:imbed}
\end{eqnarray}
\end{widetext}
where we have used the definitions
\begin{eqnarray}
s_1=\frac{\epsilon_{1i}}{{\epsilon_{1i}}^2-{\epsilon_{2i}}^2},~~~
s_2=\frac{\epsilon_{2i}}{{\epsilon_{1i}}^2-{\epsilon_{2i}}^2}.
\end{eqnarray}
The parameters $\epsilon_{1i}$ and $\epsilon_{2i}$ are defined as the values of $\epsilon_1$ and $\epsilon_2$ in the incident region, respectively.
Since the plasma frequency depends on the density, these quantities depend on the plasma density in the incident region, which we call $n_i$.
We notice that the invariant imbedding equations become singular when $\epsilon_1=0$. This singularity is at the origin of mode conversion phenomena.

The initial conditions for $r$ and $t$ are obtained from Eq.~(\ref{eq:ic}):
\begin{widetext}
\begin{eqnarray}
&&r(0)=\frac{\left({\epsilon_{1t}}^2-{\epsilon_{2t}}^2\right)\left(ip\epsilon_{1i}-q\epsilon_{2i}\right)
-\left({\epsilon_{1i}}^2-{\epsilon_{2i}}^2\right)\left(ip^\prime\epsilon_{1t}-q\epsilon_{2t}\right)}
{\left({\epsilon_{1t}}^2-{\epsilon_{2t}}^2\right)\left(ip\epsilon_{1i}+q\epsilon_{2i}\right)
+\left({\epsilon_{1i}}^2-{\epsilon_{2i}}^2\right)\left(ip^\prime\epsilon_{1t}-q\epsilon_{2t}\right)},\nonumber\\
&&t(0)=\frac{2ip\epsilon_{1i}\left({\epsilon_{1t}}^2-{\epsilon_{2t}}^2\right)}
{\left({\epsilon_{1t}}^2-{\epsilon_{2t}}^2\right)\left(ip\epsilon_{1i}+q\epsilon_{2i}\right)
+\left({\epsilon_{1i}}^2-{\epsilon_{2i}}^2\right)\left(ip^\prime\epsilon_{1t}-q\epsilon_{2t}\right)}.
\end{eqnarray}
\end{widetext}
The parameters $p$ and $q$ can be related to the incident angle $\theta$ by
\begin{eqnarray}
p=N_i k_0\cos\theta,~~
q=N_i k_0\sin\theta,
\end{eqnarray}
where $N_i$ is the effective refractive index for X waves in the incident region and is defined by
\begin{eqnarray}
N_i=\sqrt{\frac{{\epsilon_{1i}}^2-{\epsilon_{2i}}^2}{\epsilon_{1i}}}.
\label{eq:index}
\end{eqnarray}

The reflectance $R$ and the transmittance $T$ are defined by
\begin{eqnarray}
&&R=\vert r\vert^2,~~T=\frac{p^\prime}{p}\frac{\left({\epsilon_{1i}}^2-{\epsilon_{2i}}^2\right)\epsilon_{1t}}{\left({\epsilon_{1t}}^2
-{\epsilon_{2t}}^2\right)\epsilon_{1i}}\vert t\vert^2,
\end{eqnarray}
if $p^\prime$ is real. The wave becomes evanescent in the transmitted region if $p^\prime$ is imaginary. In that case, we define the transmittance to be zero.
In the absence of dissipation and mode conversion, the law of energy conservation $R+T=1$ is satisfied.
When mode conversion occurs, the absorptance $A$ ($\equiv 1-R-T$) becomes nonzero, even in the limit where the damping parameter $\nu$ goes to zero.
In that limit, $A$ represents the fraction of the incident wave energy converted to local longitudinal oscillation energy and is called as the mode conversion coefficient.

\section{Plasma density profile}
\label{sec-conf}

The invariant imbedding equations presented in the previous section can be applied to the cases where both the plasma density and the external magnetic field are {\it arbitrary} functions of the spatial coordinate $z$. In the present paper, we assume, for the sake of simplicity, that the external magnetic field is uniform in the entire space and the plasma density $n$ varies linearly such that
\begin{eqnarray}
n(z)=\begin{cases}
n_i, & \text{if } z>L,\\
n_i+n_0\frac{L-z}{\Lambda}, & \text{if } 0\le z \le L,\\
n_i+n_0\frac{L}{\Lambda}, & \text{if } z<0,
\label{eq:conf}
\end{cases}
\end{eqnarray}
where $\Lambda$ is the scale length determining the slope of the $n(z)$ curve and $n_0$ is defined by
\begin{eqnarray}
n_0=\frac{m_e \omega^2}{4\pi e^2}.
\end{eqnarray}
Waves are incident from a uniform region with $n=n_i$ onto an inhomogeneous region, and then transmitted to a uniform region with $n=n_i+(L/\Lambda)n_0$.
The upper hybrid resonance, which causes the mode conversion considered in this paper, occurs at the positions where $\epsilon_1=0$, which is equivalent
to $\omega^2={\omega_p}^2+{\omega_c}^2$. This condition can also be expressed as $X=1-Y^2$, where $X={\omega_p}^2/\omega^2$ and $Y=\omega_c/\omega$. We can rewrite $\epsilon_1$ and $\epsilon_2$ as
\begin{eqnarray}
\epsilon_1=1-\frac{X(1+i\tilde{\nu})}{(1+i\tilde{\nu})^2-Y^2},~~
\epsilon_2=\frac{X Y}{(1+i\tilde{\nu})^2-Y^2},
\end{eqnarray}
where $\tilde \nu=\nu/\omega$ and $X$ is a function of $z$.

\section{Numerical results}
\label{sec-num}

\begin{figure}
\centering\includegraphics[width=8.5cm]{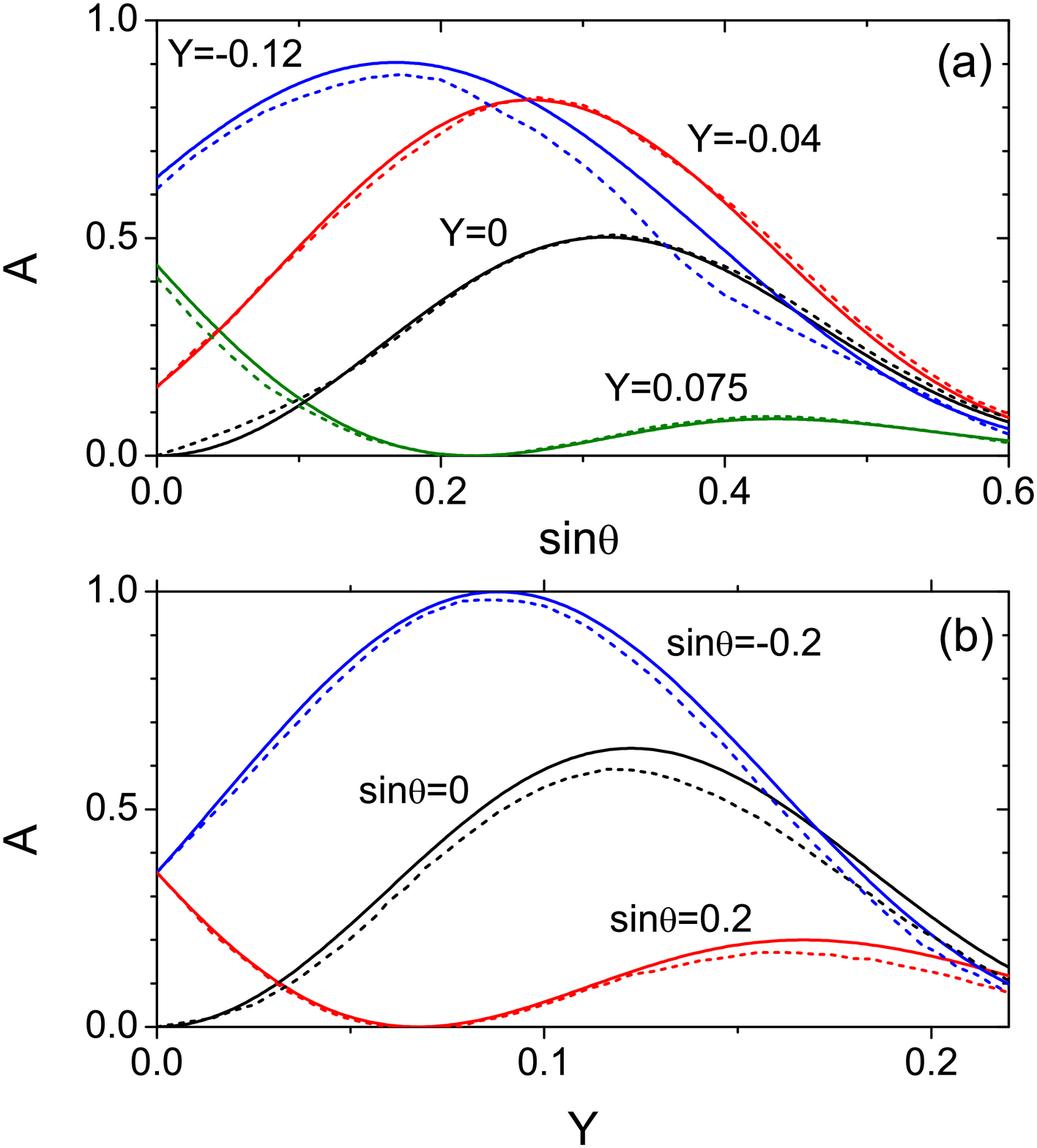}
 \caption{Mode conversion coefficient $A$ plotted (a) versus $\sin{\theta}$ when $Y ~(\equiv\omega_c/\omega) =0$, 0.075, $-0.04$, and $-0.12$
  and (b) versus $Y$ when $\sin{\theta}=0$, $0.2$, and $-0.2$. In both graphs, we have used the parameters $\zeta ~(\equiv k_0\Lambda) =10$, $\tilde{\nu}=10^{-8}$, $L/\Lambda=20$, and $n_i=0$. Our results (solid lines) are compared with those of \cite{woo} (dashed lines).}
  \label{f1}
 \end{figure}

We consider only the case where X waves are incident on an inhomogeneous plasma with the density profile given by Eq.~(\ref{eq:conf}).
Calculations in similar cases have been reported in \cite{woo,woo2,maki}. Our results are generally consistent with those of the previous works.
In Fig.~\ref{f1}, we make a comparison with the results of \cite{woo}. Though there is some discrepancy, the agreement is pretty good. We believe our results are numerically
exact solutions of the problem improving upon the previous results. In Fig.~\ref{f1}(a), the mode conversion coefficient $A$ is plotted versus $\sin{\theta}$ when $Y =0$, 0.075, $-0.04$, and $-0.12$,
while, in Fig.~\ref{f1}(b), $A$ is plotted versus $Y$ when $\sin{\theta}=0$, $0.2$, and $-0.2$.
Other parameters are fixed to $\zeta =10$, $\tilde{\nu}=10^{-8}$, and $L/\Lambda=20$. Waves are incident from a plasma-free region, therefore $n_i=0$. When $\tilde{\nu}$
is sufficiently smaller than 1, our results are independent of $\tilde{\nu}$.

\begin{figure}
\centering\includegraphics[width=8.5cm]{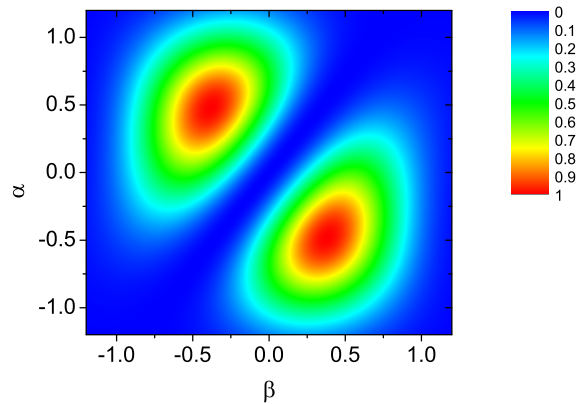}
 \caption{Color graph of the mode conversion coefficient $A$ as a function of the dimensionless parameters $\alpha=N_i\zeta^{1/3}\sin\theta$ and $\beta=\zeta^{2/3}Y$, where $N_i$ is defined by Eq.~(\ref{eq:index}). We have used the parameters $\zeta =100$, $\tilde{\nu}=10^{-8}$, $L/\Lambda=20$, and $n_i=0$, and therefore $N_i=1$. }
 \label{f2}
\end{figure}

To see the behavior of $A$ in extended parameter ranges, we show in Fig.~\ref{f2} the color graph of $A$ as a function of the dimensionless parameters $\alpha=N_i\zeta^{1/3}\sin\theta$ and $\beta=\zeta^{2/3}Y$, where $N_i$ is defined by Eq.~(\ref{eq:index}). These parameters appear naturally in the description of magnetized plasmas, as has been explained in \cite{kk2}. $\alpha$ depends on the incident angle and $\beta$ is proportional to the external magnetic field. If the
incident angle $\theta$ is changed to $-\theta$ or the direction of the external magnetic field is changed to the opposite direction,
the sign of $\alpha$ or $\beta$ is changed. In the present calculation, we have
fixed other parameters to $\zeta =100$, $\tilde{\nu}=10^{-8}$, $L/\Lambda=20$, and $n_i=0$ (therefore, $N_i=1$). The system we are considering lacks the time-reversal symmetry
due to the external magnetic field. Because of this, the wave propagation in this system is nonreciprocal and there exists an
asymmetry of physical quantities with respect to the sign change of the incident angle or $\alpha$, which is clearly manifested in Fig.~\ref{f2}. Similarly, we confirm that there is an
asymmetry with respect to the sign change of the external magnetic field or $\beta$.
In addition, we notice that there exist rather wide parameter regions where the mode conversion coefficient is greater than 0.5.
In the regions of red color, $A$ is close to 1 and the mode conversion is almost perfect.
The identification of the parameter region with strong mode conversion in various configurations
can be quite useful in the study of plasma heating phenomena both in laboratories and space.

\begin{figure}
\centering
\includegraphics[width=8.5cm]{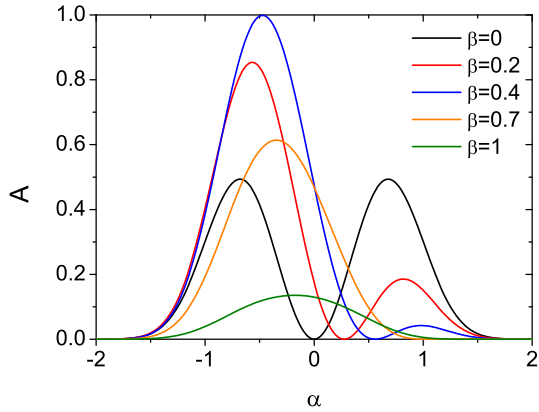}
\caption{Mode conversion coefficient $A$ plotted versus $\alpha$ when $\zeta=100$, $\tilde\nu=10^{-8}$, $L/\Lambda=20$, $n_i=0$, and $\beta=0$, 0.2, 0.4, 0.7, 1.}
\label{f3}
\end{figure}

In Fig.~\ref{f3}, we illustrate the asymmetry of the mode conversion coefficient under the sign change of the incident angle.
$A$ is plotted versus $\alpha$ when $\zeta=100$, $\tilde\nu=10^{-8}$, $L/\Lambda=20$, $n_i=0$, and $\beta=0$, 0.2, 0.4, 0.7, 1.
When there is no magnetic field ($\beta=0$), $A$ is symmetric with respect to $\alpha=0$. As the magnitude of the magnetic field increases,
the asymmetry grows rapidly. For $\beta=0.4$, we observe that the peak value of $A$ at $\alpha\approx -0.47$ is equal to 1 and perfect mode conversion occurs, whereas
the peak value at $\alpha\approx 0.98$ is only about 0.042. This strong asymmetry is in contrast with the two symmetric peaks with $A\approx 0.494$ at $\alpha\approx\pm 0.68$ when $\beta=0$.
As $\beta$ increases further, the overall magnitude of $A$ decreases gradually toward zero.

\begin{figure}
\centering
\includegraphics[width=8.5cm]{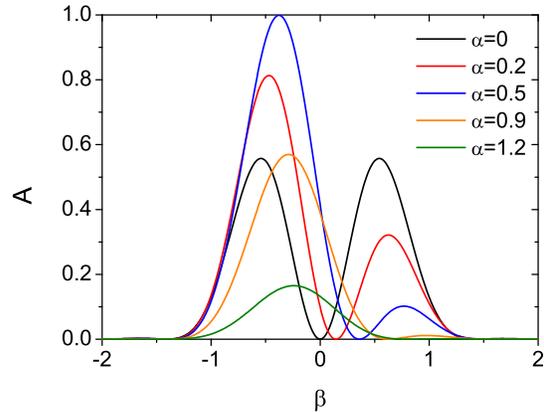}
\caption{Mode conversion coefficient $A$ plotted versus $\beta$ when $\zeta=100$, $\tilde\nu=10^{-8}$, $L/\Lambda=20$, $n_i=0$, and $\alpha=0$, 0.2, 0.5, 0.9, 1.2.}
\label{f4}
\end{figure}

In Fig.~\ref{f4}, we consider the asymmetry of the mode conversion coefficient under the sign change of $\beta$. The overall behavior is similar to the
previous case.
When the incident angle is zero (therefore, $\alpha=0$), $A$ has two peaks symmetric with respect to $\beta=0$. As $\alpha$ increases, the peak at $\beta<0$ grows
and that at $\beta>0$ decays rapidly. As $\alpha$ increases above 0.5, the overall magnitude of $A$ decreases toward zero.

\begin{figure}
\centering
\includegraphics[width=8.5cm]{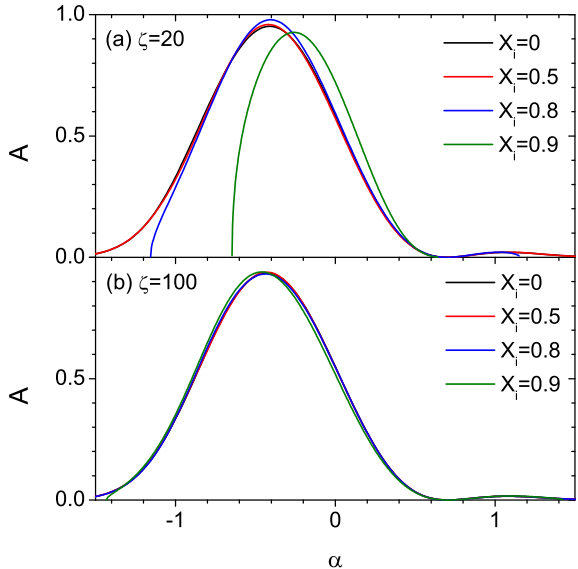}
\caption{Mode conversion coefficient $A$ plotted versus $\alpha$ (a) when $\zeta=20$ and $X_i=0$, 0.5, 0.8, 0.9 and (b) when $\zeta=100$ and $X_i=0$, 0.5, 0.8, 0.9. $X_i$ is the value of $X$ in the incident region and is proportional to $n_i$. We have used the parameters $\beta=0.5$, $\tilde\nu=10^{-8}$, and $L/\Lambda=20$.}
\label{f5}
\end{figure}

\begin{figure}
\centering
\includegraphics[width=8.5cm]{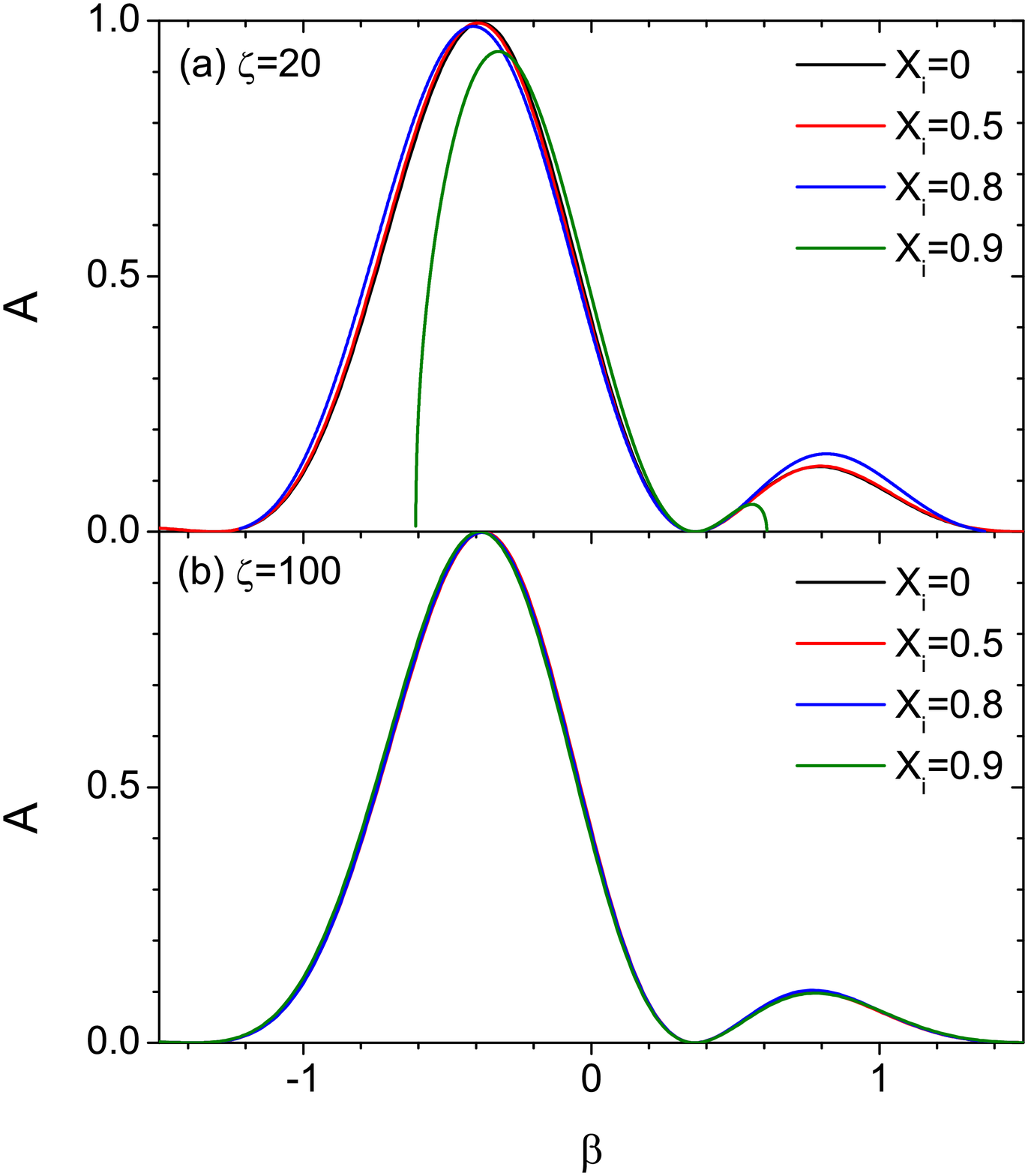}
\caption{Mode conversion coefficient $A$ plotted versus $\beta$ (a) when $\zeta=20$ and $X_i=0$, 0.5, 0.8, 0.9 and (b) when $\zeta=100$ and $X_i=0$, 0.5, 0.8, 0.9. We have used the parameters $\alpha=0.5$, $\tilde\nu=10^{-8}$, and $L/\Lambda=20$.}
\label{f6}
\end{figure}

Next we consider the dependence of the mode conversion on the plasma density in the incident region. Waves can be incident on an inhomogeneous plasma slab from a plasma-free region
where $n_i=0$ or from a uniform plasma with nonzero $n_i$. The plasma density in the incident region can be represented by the dimensionless variable $X_i$,
which is the value of $X$ in the incident region and is proportional to $n_i$. In order for an X wave to propagate and be converted into local oscillations,
the effective refractive index in the incident region should be real and the absolute value of the parameter $\alpha$ should be smaller than $\zeta^{1/3}N_i$.
For $\zeta=20$, $\beta=0.5$, and $X_i=0.9$, it is easy to show that this gives the condition that the mode conversion occurs only in the range $-0.6455<\alpha<0.6455$. Similarly, we obtain $\vert\alpha\vert<1.155$ for $X_i=0.8$ and $\vert\alpha\vert<1.91$ for $X_i=0.5$. These results are fully confirmed in Fig.~\ref{f5}(a).
For larger values of $\zeta$, the mode conversion occurs in wider ranges of $\alpha$
and the dependence of $A$ on $X_i$ becomes very weak as can be seen in Figs.~\ref{f5}(a) and \ref{f5}(b).
Equivalently, we can say that the dependence on the plasma density in the incident region becomes very weak
if the thickness of the inhomogeneous slab is sufficiently larger than the wavelength.

Finally, in Fig.~\ref{f6}, we consider the dependence of the mode conversion on the plasma density in the incident region by plotting $A$ versus $\beta$ for a fixed value of $\alpha$ ($=0.5$) and for $\zeta=20$ and 100. Similarly to Fig.~\ref{f5}, a strong dependence on $X_i$ occurs only when $X_i$ is sufficiently large and $\zeta$ is sufficiently small.
It is straightforward to show that, when $\zeta=20$ and $\alpha=0.5$, the mode conversion occurs only in the range $\vert\beta\vert<0.609$ for $X_i=0.9$
and in $\vert\beta\vert<1.366$ for $X_i=0.8$, as is confirmed in Fig.~\ref{f6}(a).

\section{Conclusion}
\label{sec-con}

In this paper, we have studied theoretically the mode conversion of electromagnetic waves into longitudinal oscillations in cold, magnetized, and stratified plasmas in the case where the external magnetic field is perpendicular to both the directions of stratification and wave propagation.
We have developed an efficient method for calculating the mode conversion coefficient for arbitrary spatial configurations of the plasma density and the external
magnetic field in a numerically precise manner using the invariant imbedding method.
We have calculated the mode conversion coefficient extensively
as a function of the incident angle, the strength of the magnetic field, and the plasma density in the incident region and
compared the results with those of previous works. We have found the parameter region where the mode conversion is almost perfect and
clarified the asymmetry of the mode conversion coefficient under the sign change of the incident angle and the external magnetic field.
Mode conversion is an important mechanism for fusion plasma heating and various astrophysical and space physics processes and our theory will be useful in the investigation of
such phenomena. In future works, we will extend our theory to the most general case where the external magnetic field is directed arbitrarily with respect to the directions
of stratification and wave propagation. We will also consider the case where both the plasma density and the external magnetic field are inhomogeneous.

\appendix
\section{Derivation of the invariant imbedding equations, Eq.~(\ref{eq:imbed})}
\label{sec-imbed}

A detailed derivation of the invariant imbedding equations in very general cases has been presented in \cite{sk}.
Here we give a brief summary of the main results.
We consider a boundary value problem of an arbitrary number of coupled first-order ordinary differential equations defined by
\begin{eqnarray}
&&\frac{d}{dz} {\bf u}(z)={\bf F}(z,{\bf u}(z)),~~ z \in [0,L],\label{eq:uu} \\
&&g{\bf u}(0)+h{\bf u}(L)={\bf v},\label{eq:bc}
\end{eqnarray}
where $\bf u$, $\bf F$, and $\bf v$ are $N$-component vectors and $g$ and $h$ are $N\times N$ matrices.
We consider the function $\bf u$ as being dependent on $L$ and $\bf v$:
\begin{eqnarray}
{\bf u}(z)={\bf u}(z;L,{\bf v})
\end{eqnarray}
and define
\begin{eqnarray}
{\bf R}(L,{\bf v})={\bf u}(L;L,{\bf v}),~~~ {\bf S}(L,{\bf v})={\bf u}(0;L,{\bf v}).
\end{eqnarray}
Though the general method presented in \cite{sk} can be applied to nonlinear cases,
we restrict to the case where ${\bf F}(z,{\bf u})$ is linear in $\bf u$ such that
\begin{eqnarray}
F_i(z,{\bf u}(z))=A_{ij}(z)u_j(z).
\end{eqnarray}
Then the functions ${\bf u}$, $\bf R$, and $\bf S$ are linear in $\bf v$:
\begin{eqnarray}
&&{\bf u}(z;L,{\bf v})=U(z;L){\bf v},\nonumber\\&&
{\bf R}(L,{\bf v})=R(L){\bf v},~~{\bf S}(L,{\bf v})=S(L){\bf v},
\label{eq:urs}
\end{eqnarray}
where $U$, $R$, and $S$ are $N\times N$ matrices.
After some considerations, it is possible to derive
the invariant imbedding equations satisfied by $U$, $R$, and $S$. Here we only give the equations for $R$ and $S$:
\begin{eqnarray}
\frac{d}{dl}R(l)&=&A(l)R(l)-R(l)hA(l)R(l),\nonumber\\
\frac{d}{dl}S(l)&=&-S(l)hA(l)R(l),
\label{eq:imbed1}
\end{eqnarray}
which satisfy the initial conditions
\begin{eqnarray}
R(0)=S(0)=(g+h)^{-1}.
\label{eq:ic}
\end{eqnarray}

We now apply the invariant imbedding method to Eq.~(\ref{eq:meq}).
We are mainly interested in calculating the reflection and transmission coefficients $r$ and $t$ defined by Eq.~(\ref{eq:psi}).
We define the values of $\epsilon_1$ and $\epsilon_2$ in the incident region ($z>L$) as $\epsilon_{1i}$ and $\epsilon_{2i}$
and those in the transmitted region ($z<0$) as $\epsilon_{1t}$ and $\epsilon_{2t}$, respectively.
At the boundaries of the inhomogeneous medium, we have
\begin{eqnarray}
u_1(0;L)&=&t(L),~~~u_1(L;L)=1+r(L),\nonumber\\
u_2(0;L)&=&\frac{1}{{\epsilon_{1t}}^2-{\epsilon_{2t}}^2}\left(-ip^\prime\epsilon_{1t}+q\epsilon_{2t}\right)t(L)\nonumber\\
&=&\frac{1}{{\epsilon_{1t}}^2-{\epsilon_{2t}}^2}\left(-ip^\prime\epsilon_{1t}+q\epsilon_{2t}\right)u_1(0;L),\nonumber\\
u_2(L;L)&=&\frac{1}{{\epsilon_{1i}}^2-{\epsilon_{2i}}^2}\left\{ip\epsilon_{1i}\left[r(L)-1\right]+q\epsilon_{2i}\left[r(L)+1\right]\right\}\nonumber\\
&=&\frac{1}{{\epsilon_{1i}}^2-{\epsilon_{2i}}^2}\left(ip\epsilon_{1i}
+q\epsilon_{2i}\right)u_1(L;L)\nonumber\\
&&-2ip\frac{\epsilon_{1i}}{{\epsilon_{1i}}^2-{\epsilon_{2i}}^2}.
\label{eq:bcw}
\end{eqnarray}
From Eq.~(\ref{eq:bcw}), we have
\begin{eqnarray}
g{\bf{S}}+h{\bf{R}}={\bf v},
\end{eqnarray}
where
\begin{eqnarray}
&&{\bf{S}}=\begin{pmatrix}u_1(0;L)\\u_2(0;L)\end{pmatrix},~~{\bf{R}}=\begin{pmatrix}u_1(L;L)\\u_2(L;L)\end{pmatrix},\nonumber\\
&&{\bf v}=\begin{pmatrix} 0\\v_2\end{pmatrix}=\begin{pmatrix} 0\\2ip\frac{\epsilon_{1i}}{{\epsilon_{1i}}^2-{\epsilon_{2i}}^2}\end{pmatrix}
,\nonumber\\
&&g=\begin{pmatrix} \frac{1}{{\epsilon_{1t}}^2-{\epsilon_{2t}}^2}\left(ip^\prime\epsilon_{1t}-q\epsilon_{2t}\right) & 1\\ 0&0\end{pmatrix},\nonumber\\
&&h=\begin{pmatrix} 0 & 0\\ \frac{1}{{\epsilon_{1i}}^2-{\epsilon_{2i}}^2}\left(ip\epsilon_{1i}
+q\epsilon_{2i}\right) &-1\end{pmatrix}.
\label{eq:dh}
\end{eqnarray}
From the definitions of $R$ and $S$ given by Eq.~(\ref{eq:urs}),
we obtain
\begin{eqnarray}
&&R_{12}v_2=1+r(L),~~S_{12}v_2=t(L),\nonumber\\
&&R_{22}v_2=
\frac{1}{{\epsilon_{1i}}^2-{\epsilon_{2i}}^2}\left\{ip\epsilon_{1i}\left[r(L)-1\right]\right.\nonumber\\&&
\left. ~~~~~~~~~~~+q\epsilon_{2i}\left[r(L)+1\right]\right\}.
\label{eq:rtd}
\end{eqnarray}
From Eqs.~(\ref{eq:rtd}) and (\ref{eq:imbed1}), in which the expressions for $A$, $h$, and $v_2$ given by Eqs.~(\ref{eq:da}) and ({\ref{eq:dh}) are used,
we derive the invariant imbedding equations for $r$ and $t$, Eq.~(\ref{eq:imbed}), in a straightforward manner.

\acknowledgments
This work has been supported by the National Research Foundation of Korea Grant (NRF-2020R1A2C1007655) funded by the Korean Government.
S.K. has also been supported by the Basic Science Research Program through the National Research Foundation of Korea Grant (NRF-2020R1A6A3A01098816) funded by the Ministry of Education.

\end{document}